\documentclass{article}

\usepackage{graphicx}
\usepackage{amsmath}
\usepackage{amssymb}
\usepackage[preprint]{neurips_2025}

\usepackage[utf8]{inputenc} 
\usepackage[T1]{fontenc}    
\usepackage{hyperref}       
\usepackage{url}            
\usepackage{booktabs}       
\usepackage{amsfonts}       
\usepackage{nicefrac}       
\usepackage{microtype}      
\usepackage{xcolor}         

\usepackage{subcaption}  
\usepackage{algorithm}
\usepackage{algorithmic}
\usepackage{siunitx}

\title{BeamClean: Language Aware Embedding Reconstruction}

\author{%
  Kaan Kale  \\
  Protopia AI, \\
  Austin, TX, USA \\
  \texttt{kaan.kale@protopia.ai} \\
  \And
    Kyle Mylonakis \\
    Protopia AI, \\
    Austin, TX, USA \\
    \texttt{kyle@protopia.ai} \\
  \AND
    Jay Roberts \\
    Protopia AI, \\
    Austin, TX, USA \\
    \texttt{jay@protopia.ai} 
  \And
    Sidhartha Roy \\
    Protopia AI, \\
    Austin, TX, USA \\
    \texttt{sid@protopia.ai} \\
}

\begin{document}

\newcommand{\langmodel}{\ifmmode p_{\mathrm{LM}} \else Prior Model\fi}
\newcommand{\staticnoise}{input-independent noise}
\newcommand{\servicetype}{MaaS}
\newcommand{\approach}{\texttt{BeamClean}}
\newcommand{\surrogate}[1][]{\ensuremath{\pi_{\hat{\theta}^{#1}}}}

\maketitle

\begin{abstract}
  In this work, we consider an inversion attack on the obfuscated input embeddings sent to a language model on a server, where the adversary has no access to the language model or the obfuscation mechanism and sees only the obfuscated embeddings along with the model’s embedding table. We propose \approach{}, an inversion attack that jointly estimates the noise parameters and decodes token sequences by integrating a language-model prior. Against Laplacian and Gaussian obfuscation mechanisms, \approach{} always surpasses naive distance-based attacks. This work highlights the necessity for and robustness of more advanced learned, input-dependent methods.
\end{abstract}

\section{Introduction}
Machine learning services increasingly rely on shared or outsourced resources to manage and process data. Model-as-a-Service (MaaS) is a prime example, where organizations outsource the generation and storage of pre-trained large language models, computing infrastructure, and core functionalities that enable users to fine-tune, deploy, and execute customized models.

While this arrangement offers scalability and cost benefits, it also heightens privacy risks, particularly for Large Language Models (LLMs). Plaintext data must be transmitted from the client's trust zone to enter the model provider's. When the data is at rest in the client's trust zone or in transit to the model provider, the plaintext data may be protected via encryption, however, while the data is in use by an LLM, the data must be converted to either plaintext tokens or word embeddings (which are in 1-1 correspondence with the plaintext tokens). This need to expose data while it is in use presents a heightened privacy risk as this plaintext data could be leaked or intercepted.

\begin{figure}[h]
  \centering
\includegraphics[width=1\textwidth, trim=0mm 98mm 6mm 29mm, 
    clip]{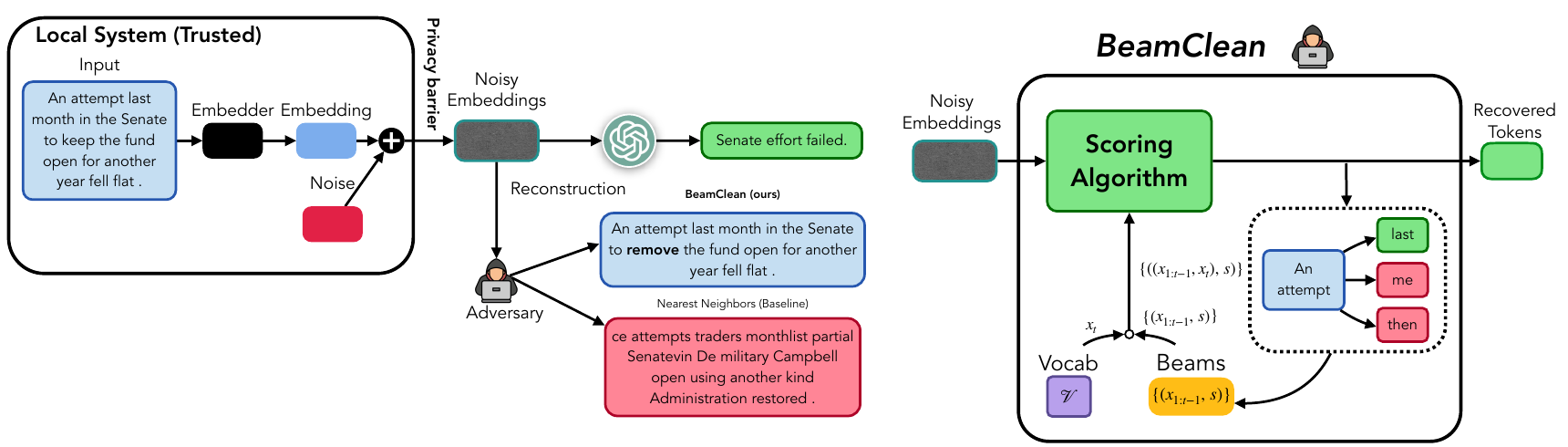}  
  \caption{Overview of the generic input-embedding obfuscation pipeline and our adversarial threat model. Plaintext inputs are first encoded and transformed into noisy (i.e. obfuscated) embeddings, which are then transmitted to the LLM provider. An attacker accesses the noisy embeddings in order to attempt to recover the original plaintext input. Within the local trust zone, an obfuscation mechanism is applied to embeddings for a target LLM. These noisy embeddings are inputs to the \approach{} algorithm.  The noisy embeddings are then put through a scoring algorithm to determine the top-k candidate token-ids. These top candidates are added to candidates from previously scored tokens to form beams. The top scoring beams are selected and used to start the scoring algorithm for the next token in the sequence.
  }
  \label{fig:pipeline_fig_1}
\end{figure}

To mitigate these privacy risks, recent approaches introduce input-independent noise to token embeddings in the local trust zone as in, Fig~\ref{fig:pipeline_fig_1}, before sharing with the service provider. One common method is local differential privacy techniques \cite{mai2023split,shen2023split,feyisetan-kasiviswanathan-2021-private,carvalho2023tem} which add input independent noise to the embeddings. This paper investigates the possibility of inverting input-independent noise‑perturbed input word embeddings back to their original token counterparts to recover the plaintext data.

\begin{figure}[t]
  \centering
\includegraphics[width=\textwidth, trim=22mm 130mm 127mm 39mm, 
    clip]{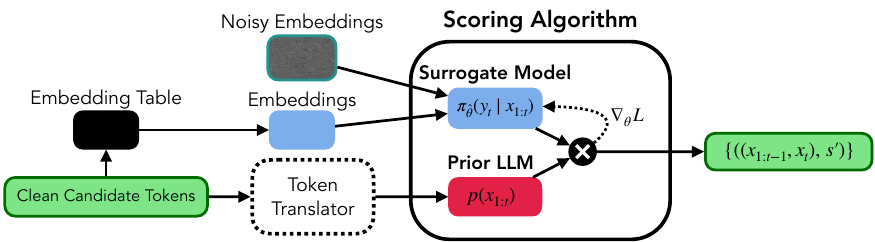} \caption{
\approach{} is an iterative algorithm that begins with clean candidate tokens mapping to their corresponding embeddings. These clean candidate embeddings and noisy embeddings are inputs to a surrogate noise model of the obfuscation mechanism, $\surrogate[]$. The clean candidate tokens are also used to produce a language prior (optionally translating tokens for the case of distinct target and prior language models). Together, the language prior and the surrogate noise model produce a likelihood score which is used to train the surrogate model. This is done iteratively to update the beam candidates. Finally, the highest scoring beam is selected as the reconstruction.
}
  \label{fig:pipeline_fig}
\end{figure}

We propose a novel attack strategy, \approach{} \footnote{\url{https://github.com/beamclean-neurips25/beamclean}}
(Fig \ref{fig:pipeline_fig}), aimed at reconstructing sequences of tokens from their obfuscated embeddings. We assume a scenario, wherein the adversary has no access to the target model’s internal parameters, the obfuscation mechanism, or its training data. Instead, the attacker only observes the obfuscated embeddings \citet{mai2023split} and has knowledge of the underlying embedding table. This attack scenario could occur if an unauthorized user gains access to the server's compute infrastructure.

Our results demonstrate that \approach{} (Fig. ~\ref{fig:pipeline_fig}) substantially outperforms nearest neighbor-based methods previously used to assess privacy strength \cite{mai2023split} against noise mechanisms that do not vary with input, effectively reconstructing plaintext sequences. This emphasizes a critical insight: Input independent Gaussian and Laplacian noise mechanisms, like those used in local-DP, can be vulnerable against adversaries equipped with linguistic priors and embedding knowledge.

The structure of this paper is as follows. Section \ref{sec:related_work} reviews the related work. Section \ref{sec:formulation} introduces the problem setting under consideration. In Section \ref{sec:methodology}, we develop the framework and propose our attack method. Next, Section \ref{sec:experiments} presents experimental results to demonstrate the effectiveness of our proposed approach. Finally, Section \ref{sec:conlusion} provides concluding remarks.

\section{Related Work}
\label{sec:related_work}
With the growth of large language models (LLMs), chatbots such as ChatGPT, LLama \cite{touvron2023llama}, \cite{achiam2023gpt}, and other embedding model services have raised significant privacy concerns. These services allow for sensitive or proprietary data to be transmitted during both training and inference, potentially leading to privacy risks.
The main focus is to protect against these privacy risks without sacrificing model performance. Existing research primarily focuses on centralized learning and the leakage of training data in public LLM deployments, with attention given to pre-training \cite{li2021large}, fine-tuning \cite{yu2021differentially}, and prompt-tuning \cite{li2023privacy}. However, little work has addressed local privacy\cite{chatzikokolakis2013broadening} during the inference phase.

\paragraph{Text Reconstruction from Contextualized Embeddings}  
Previous studies have focused on inverting the contextualized embeddings produced by an \emph{embedding model} \cite{kugler2021invbert, morris2023textembeddingsrevealalmost}. These studies demonstrate that high BLEU scores can be achieved when an attacker accesses the model's output. In \cite{morris2023textembeddingsrevealalmost}, the authors introduced Gaussian noise as a simple obfuscation technique and presented preliminary results showing that, at a modest noise level, reconstruction accuracy dropped drastically (BLEU score fell from over 80 to around 10), while retrieval performance was barely affected. These findings suggest that noise injection could be a practical way to protect sensitive text data stored in vector databases without sacrificing search utility.

\paragraph{Input Data Protection in Split Learning}
In \cite{mai2023split}, the authors propose a method for protecting input data in split learning by adding Laplacian noise under a local differential privacy framework. Their approach includes a pipeline that post-processes the contextualized embeddings generated by an embedding model. In \cite{shen2023split}, a similar strategy is adopted with an additional step: after adding Laplacian noise, the noisy embedding is mapped to its nearest neighbor, resulting in a change of the corresponding input token.


\paragraph{Word Embedding Perturbation Mechanisms}
In \cite{carvalho2023tem}, the authors propose a truncated exponential mechanism that, rather than directly perturbing the continuous word embedding, assigns each candidate word a score based on its negative distance from the input, adds calibrated Gumbel noise to these scores, and selects the word with the highest noisy score as the privatized output. This method dynamically adapts the noise to the local density of the embedding space, ensuring that the selected token remains close to the original while providing formal privacy guarantees. In contrast, \cite{xu2020differentially} replaces standard spherical noise with elliptical noise sampled from a density proportional to 
$\exp\left(-\epsilon \|z\|_{\text{RM}}\right)$,  where the regularized Mahalanobis norm adjusts the noise according to the covariance structure of the embedding space. The perturbed embedding is then mapped to its nearest token, thereby enhancing privacy in sparse regions without sacrificing overall utility. Meanwhile, \cite{feyisetan2020privacy} presents a mechanism under $d_\chi$-privacy that maps each word into a high-dimensional embedding space, adds noise sampled from an $n$-dimensional distribution with density proportional to $\exp\left(-\epsilon \|z\|\right)$, and then maps the perturbed vector to its nearest neighbor in the vocabulary. 

All three methods share the common step of adding input-independent noise to embeddings and then mapping back to the nearest token. Their key differences lie in the noise model and privacy framework: \cite{carvalho2023tem} and \cite{xu2020differentially} operate under standard differential privacy \cite{xu2020differentially} adopt noise based on local geometry—whereas \cite{feyisetan2020privacy} adopts $d_\chi$-privacy, which defines privacy with respect to a metric over the embedding space.

\section{Problem Formulation}
\label{sec:formulation}
We consider a setting in which an attacker gains access to obfuscated embeddings derived from sensitive text, but lacks direct knowledge of how these embeddings were transformed. In this section, we formalize the obfuscation mechanism, define the blind attack scenario, and state the inversion problem underpinning our proposed attack strategy. Though the obfuscation mechanisms tested in this paper are input-independent, \approach{} is mathematically formulated for the input and sequence dependent cases.

We begin by outlining some useful notation to be used throughout the rest of the paper. Let $\mathcal{V} = \{w_1, \dots, w_{|\mathcal{V}|}\}$ denote the vocabulary of the target language model, and let each vocabulary word $w_i \in \mathcal{V}$ have an associated clean embedding $x_i \in \mathbb{R}^d$ from the embedding table $\mathcal{X}$. 

Given a token sequence $w_{1:T} = (w_1, \dots, w_T)$ of length $T$, where each $w_t \in \mathcal{V}$ for $t \in [T] := \{1, \dots, T\}$, we define $x_t := x(w_t)$ to be the embedding corresponding to token $w_t$. Thus, the sequence of embeddings is denoted by $x_{1:T} = (x_1, \dots, x_T)$. Also, for notational simplicity, we use the expressions $
  p_{\theta}(x)
  \quad\text{and}\quad
  p(x;\theta)$
interchangeably.

The \emph{target language model} is the cloud‑hosted LLM which processes obfuscated embeddings of users' plaintext data. An attacker obtains these obfuscated embeddings and has access only to that model’s embedding table and vocabulary. An \emph{obfuscation mechanism} is a map $\mathcal{O}(\cdot; \theta)$ between a clean embedding $x_t$ to an \emph{obfuscated} embedding $y_t$ with $t \in [T]$, parameterized by $\theta$: $y_{1:T} \;=\; \mathcal{O}\bigl(x_{1:T}; \theta\bigr) 
    \;=\; \bigl(y_1, y_2, \dots, y_T\bigr).
$

In this work, we focus on \emph{additive-noise} mechanisms where $
y_t \;=\; x_t +n_t, \forall t \in [T]$ and each noise term $n_t$ is drawn from a distribution ${p\bigl(n_t \mid x_{1:T}; \theta\bigr)}$. 

The noise is similar to the noise mechanisms used in differential privacy. We consider an \textbf{ attack scenario}, in which the adversary has access only to: The \emph{obfuscated embeddings} $y_{1:T}$ (leaked from some system). The \emph{embedding table} $\mathcal{X}$ of the target language model, which maps each token $w_i \in \mathcal{V}$ to its clean embedding $x_i$. The attacker does not have the target model’s internal parameters (i.e. the model weights or model architecture), nor knowledge of the obfuscation mechanism $\mathcal{O}(\cdot; \theta)$, nor its training data. Such a scenario may arise when only partial leaks reveal the initial embedding layer and the obfuscated inputs, but no deeper components. This situation naturally arises when multiple LoRa finetuned adapters, which typically leave the embedding layer as it is, are associated with the same base model via inference engines such as vLLM \cite{kwon2023efficient}.

We make the following novel contributions to the inversion of transformed token embeddings to text:

\paragraph{Generalized Attack Framework.} We introduce a novel approach for attacks against additive-noise obfuscation methods on sequential data. Even without direct access to the obfuscation mechanism or its parameters, our method adapts to both input-independent and input-adaptive noise. 

\paragraph{Improved Inversion over Nearest-Neighbor.}
For sequences of word embeddings, \approach{} outperforms nearest-neighbor based reconstruction attacks against \staticnoise{} obfuscations, similar to those used in local-DP \cite{feyisetan-kasiviswanathan-2021-private}.

\paragraph{Language Aware Reconstruction.}  Our method performs denoising of token embedding sequences by modeling them as interdependent, rather than treating individual tokens in isolation. This approach enables the incorporation of sequential dependencies present in language as priors during reconstruction, leveraging a pretrained language model.

\section{BeamClean}
\label{sec:methodology} 
The core idea of \approach{} is to jointly estimate the noise model parameters and decode the original token sequence using a beam-search based procedure that leverages both the embedding table $\mathcal{X}$ of the target model and a language prior.

Let $p(y \mid x; \theta)$ denote the noise model—i.e., the probability of observing the noisy embedding $y$ given its clean counterpart $x$. Our goal is to obtain a maximum-a-posteriori (MAP) estimate of the noise-model parameters $\theta$ conditioned on the observed sequence $y_{1:T}.$ Applying Bayes’ rule and marginalizing over the unknown clean token sequence $w_{1:T}$ yields the following objective:

\begin{equation}
\label{eq:theta_est}
  \hat{\theta} =  \arg\max_{\theta}\,\log p(\theta \mid y_{1:T}) = \arg\max_{\theta}\,\
    \log\sum_{w_{1:T}\in\mathcal V^T}\pi_\theta\bigl(y_{1:T}\mid x(w_{1:T}))\,
           p_{\mathrm{LM}}(w_{1:T}) + \log p(\theta)
\end{equation}

In this formulation, because we assume an uninformative uniform prior, the $p(\theta)$ term is constant and drops out of the optimization. A full expression and detailed derivation are provided in Appendix~\ref{app:sequential}. The key terms in Equation~\eqref{eq:theta_est} are as follows:
\begin{itemize}
    \item $\pi_{\theta}$, is the surrogate noise model. The likelihood term \(\pi_{\theta}(y_{1:T} \mid x_{1:T})\) is factorized as:
    $\pi_\theta(y_{1:T} \mid x_{1:T}) \triangleq \prod_{t=1}^{T} \pi_\theta(y_t \mid x_{1:T})$. If there is more information available about the obfuscation mechanism it can be incorporated to the surrogate noise model by parametrizing it accordingly. For example if we know that the obfuscation mechanism is Gaussian we can model it as:
    \begin{equation}
    \pi_{\theta}(x(w_{1:t}))
        \;=\;
        \mathcal{N}
        \bigl(
            y_t \,\big|\,
            x_t + \mu_{\theta}(x_{1:t}),\, \Sigma_\theta(x_{1:t})
        \bigr),
        \label{eq:gauss_surrogate}
    \end{equation}
    
    where $\mu_{\hat{\theta}}(\cdot)$ and $\Sigma_{\hat{\theta}}(\cdot)$ are the mean and covariance predicted by the surrogate noise model, respectively, for each time step $t$.

    \item $\langmodel(w_{1:T})$ is the prior language model over the token sequence, defined as $\langmodel(w_{1:T}) \;=\; \prod_{t=1}^{T} \langmodel\!\bigl(w_t \mid w_{1:t-1}\bigr)$ and it assigns low probability to linguistically implausible sequences.
\end{itemize}

Once we have an estimate of the parameters $\hat{\theta}$, we can decode the token sequence by maximizing the posterior of the tokens $w_{1:T}$ given the observed transformed embedding $y_{1:T}$:
\begin{equation}
\label{eq:decode_formal}
\begin{aligned}
    \hat{w}_{1:T}
    = \arg\max_{w_{1:T}}
       \log p(w_{1:T} \mid y_{1:T}; \hat{\theta}) = \arg\max_{w_{1:T}}
       \log
           \pi_{\hat{\theta}}\bigl(y_{1:T}\mid x(w_{1:T})\bigr)\,
           p_{\mathrm{LM}}(w_{1:T})
\end{aligned}
\end{equation}

Evaluating the sum in Equation~\eqref{eq:theta_est} requires enumerating all
$|\mathcal V|^{\,T}$ token sequences, which is prohibitive for realistic
vocabulary sizes and sequence lengths.  
To achieve tractable computation, we rely on two complementary approximations:

\begin{enumerate}
    \item \textbf{Causal noise model.}
   We assume the obfuscation mechanism is causal,
   \[
       p_{\theta}\bigl(y_t \mid w_{1:T}\bigr)
       \;=\;
       p_{\theta}\bigl(y_t \mid w_{1:t}\bigr),
       \qquad t<T,
   \]
   so each noisy embedding depends only on the current and past clean tokens.
   \item \textbf{Beam-search pruning.}
   Even with causality the number of candidate
   sequences still grows exponentially in $T$.  
   We therefore keep only the top-$B$ partial hypotheses at each time-step,
   selected by beam search (see Section~\ref{sec:decoding}).
\end{enumerate}

Combining the causal assumption with beam-search pruning reduces the overall
complexity from $\mathcal O(|\mathcal V|^{\,T})$ to
$\mathcal O(BT)$, making optimization feasible in practice.

\begin{algorithm}[ht]
    \caption{Causal Beam Search Decoding with Adaptive Noise Estimation}
    \label{alg:decoding}
    \begin{algorithmic}[1]
        \REQUIRE Vocabulary $\mathcal{X}$, beam size $k$, sequence length $T$, noisy embeddings $y_{1:T}$
        \STATE Initialize beam $\mathcal{B}_0 \gets \{(\,\texttt{""}, 1.0)\}$ \hfill $\triangleright$ Empty sequence with score 1
        \STATE Initialize noise parameters $\theta^{(0)}$ (e.g.\ randomly or via a pre-training step)
        \FOR{$t=1$ {\bfseries to} $T$}
            \STATE $\theta^{(t)} \gets \arg\max_{\theta} \sum_{(x_{1:t-1},s) \in \mathcal{B}_{t-1}} \sum_{x_t \in \mathcal{X}} \surrogate(y_t \!\mid\! x(w_{1:t})\bigr)\langmodel(w_t\mid w_{1:t-1}\bigr)$
            \STATE $\mathcal{C} \gets \varnothing$ \hfill $\triangleright$ Set of new candidates
            \FOR{\textbf{each} $(w_{1:t-1}, s)$ in $\mathcal{B}_{t-1}$}
                \FOR{\textbf{each} $w_t$ in $\mathcal{X}$}
                    \STATE $s' \gets s \;\times\; 
                    \surrogate(y_t \mid x(w_{1:t}))
                    \;\times\; \langmodel(w_t \mid w_{1:t-1})$
                    \STATE $\mathcal{C} \gets \mathcal{C} \,\cup\, \bigl\{(w_{1:t-1},w_t),\; s'\bigr\}$
                \ENDFOR
            \ENDFOR
            \STATE $\mathcal{B}_t \gets \text{Top-}k \text{ entries of }\mathcal{C}\text{ by score}$ 
        \ENDFOR
        \STATE \RETURN{The highest-scoring sequence in $\mathcal{B}_T$.}
    \end{algorithmic}
\end{algorithm}
\label{sec:decoding}
We employ a beam search approach that iteratively refines both the noise parameters $\theta$ and the decoded tokens. At each time step $t$, we keep a ``beam'', $\mathcal{B}_{t-1}$ of candidate partial sequences, each with an associated score. Lines 1–2 initialize an empty beam with the start‐of‐sequence hypothesis. At each time step $t$ (lines 3–5), the surrogate noise parameters $\theta^{(t)}$ are updated by maximizing the joint likelihood of the current noisy embedding and all beam sequences. Lines 6–9 then extend each beam hypothesis by every token in the vocabulary, computing a new score by multiplying the surrogate likelihood (Eq.) with the language‐model prior. Finally, lines 10–11 prune to the top-$B$ sequences to form the next beam, and this process repeats until $T$, with the best scoring sequence returned.

\smallskip
\noindent\textbf{Adaptive Noise Estimation.}
At each time step, re-estimating $\theta$ with the current beam $\mathcal{B}_{t-1}$ ensures that the model continually refines its approximation of the true obfuscation process. This incremental approach leverages context gained from earlier steps to improve parameter estimates.

\smallskip

\noindent\textbf{Efficient Parameter Initialization.}
Using $\theta^{(t-1)}$ as the initialization for $\theta^{(t)}$ reflects the assumption that a single noise model operates across the entire sequence. Rather than refitting from scratch at each step, we exploit continuity in the underlying noise parameters.

\smallskip
\noindent
Overall, this methodology interleaves adaptive noise-model estimation with a causal beam search for token decoding. By balancing linguistic plausibility (via a language prior) with consistency under the learned noise distribution, our approach substantially improves inversion accuracy over naive baselines, especially for \staticnoise{} mechanisms.

\section{Experiments}
\label{sec:experiments}

We evaluate \approach{} across multiple datasets and Gaussian and Laplacian noise mechanisms, similar to those used in local-DP. We analyze performance under varying levels of adversarial knowledge and noise complexity. \approach{} is compared against the common baseline in blind-obfuscation literature of \emph{Nearest Neighbor } \cite{mai2023split, Du_2023, xu2020differentially}, which decodes each noisy embedding $y_t$ to the single clean embedding $x \in \mathcal{X}$ that minimizes $\|y_t - x\|_2$. 

Our study uses three datasets. The first is constructed from randomly sampled examples in the Open-Orca dataset \cite{mukherjee2023orca}. To standardize the input length, we truncate all sequences to a maximum of 32 tokens. The second dataset is the MRPC dataset from the GLUE benchmark \cite{wang2019gluemultitaskbenchmarkanalysis}. Finally, we use the PAPILLON dataset to evaluate PII recovery rate \cite{siyan2025papillon}.%

Following our problem formulation (Section~\ref{sec:methodology}), we incorporate a pretrained Llama-3.2-1B-Instruct model as the language prior in Equation \eqref{eq:theta_est}. 

We report the \emph{attack success rate} ({ASR}), the fraction of correctly recovered tokens in a sequence:
\[
\mathrm{ASR}
\;=\;
\frac{\text{Number of Correctly Decoded Tokens}}{\text{Total Number of Tokens}} 
\times 100\%.
\]
Higher values indicate more successful reconstructions. 

For the PAPILLON benchmark, we report the mean PII recovery percent. For each sample in the dataset, we use PAPILLON to measure the percentage of PII strings leaked in the sample (based on their prompt \footnote{\url{https://github.com/Columbia-NLP-Lab/PAPILLON}}). We normalize PII recovery percent by the total number of PII strings detected in the clean sample and report the average value across the dataset.

To assess how our attacks perform under different privacy levels, we compute the corresponding differential privacy parameter $\epsilon$ for various noise magnitudes in the input-independent setting. The calculations for each noise distribution are provided in Appendix \ref{app:epsilon_calc}.

Our experiments were executed on {NVIDIA H100} GPUs with 80\,GB of memory. Training time grows approximately linearly with the beam width, dataset size, embedding‑table size, and the number of candidate tokens considered.

\subsection{Results}
\label{subsec:results}

\begin{figure}[t]
    \begin{subfigure}[t]{0.45\textwidth}
    \centering
    \includegraphics[width=\textwidth]{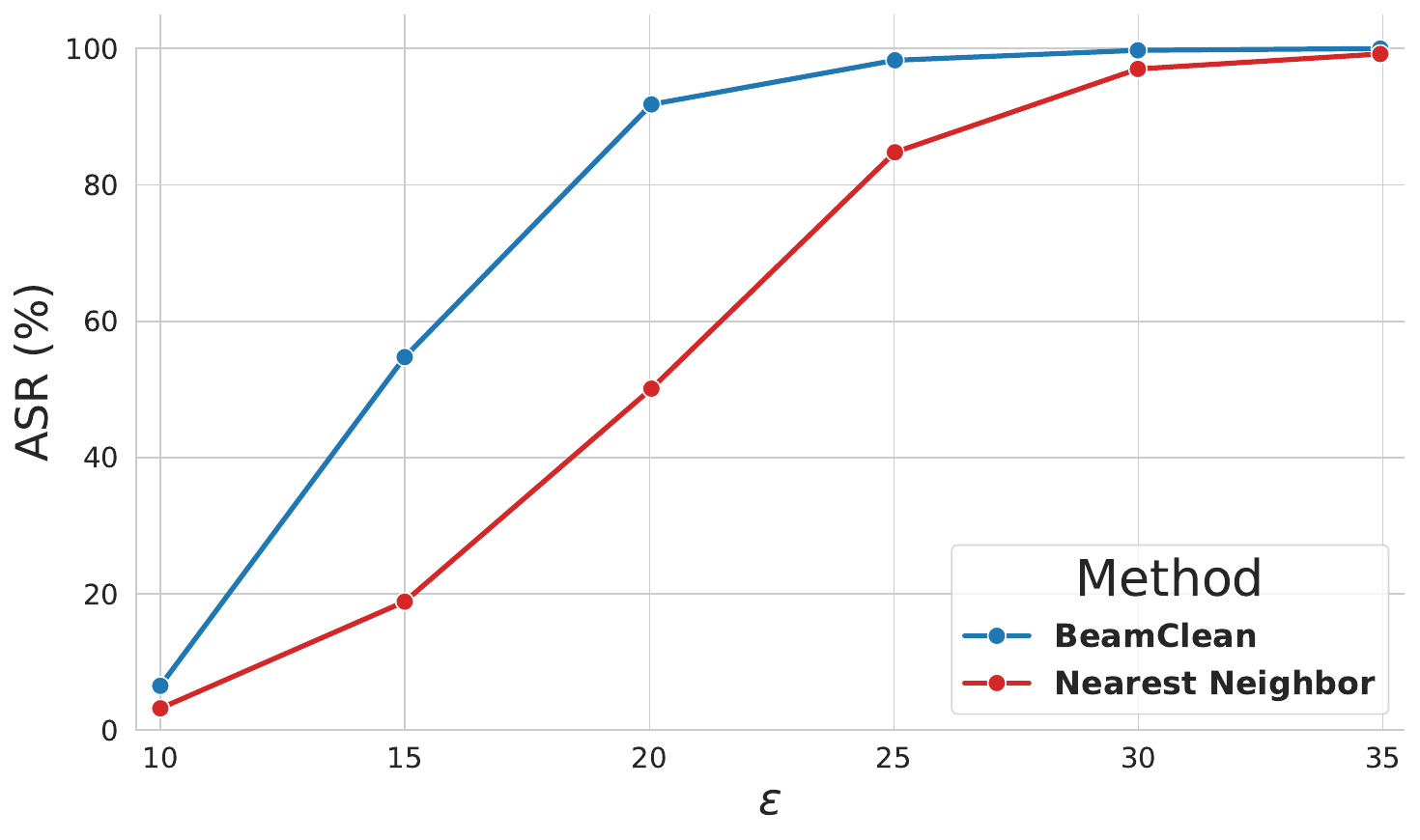}
        \caption{Gaussian noise mechanism}
        \label{fig:comparison-mrpc-g}
    \end{subfigure}
    \centering
        \hspace{0.03\textwidth} 
    \begin{subfigure}[t]{0.45\textwidth}
        \centering
        \includegraphics[width=\textwidth]{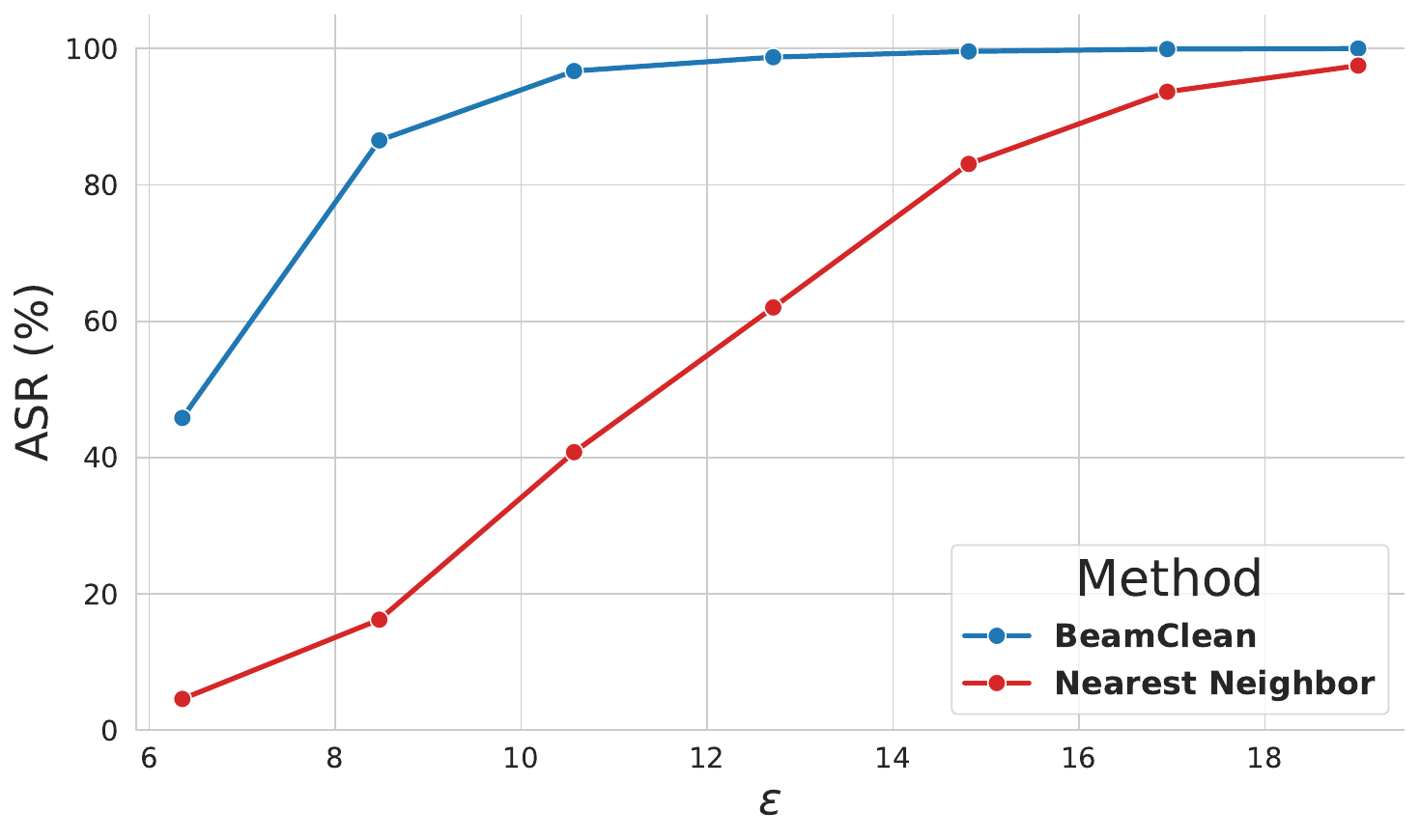}
       \caption{Laplace noise mechanism}
    \label{fig:comparison-mrpc-l}
    \end{subfigure}
    \caption{Performance of \approach{} compared to Nearest Neighbor on the MRPC dataset. Curves show token-recovery rate as a function of~$\epsilon$ with beam size~20.
    We compare against Gaussian, \ref{fig:comparison-mrpc-g}, and Laplacian, \ref{fig:comparison-mrpc-l}, noise mechanisms using, respectively.
    In both cases \approach{} outperforms Nearest Neighbor. Against Gaussian noise at $\epsilon=15$ our attack recovers {74.3\%} of tokens versus {42.1\%} for Nearest Neighbor. Against Laplacian noise at $\epsilon=8.5$ the attack attains {86\%} recovery versus {18\%} for Nearest Neighbor.
    }
    \label{fig:comparison-mrpc}
\end{figure}

\paragraph{BeamClean always outperforms Nearest Neighbor.}
\label{subsub:static_result}
Figure~\ref{fig:comparison-mrpc} presents an experiment conducted on the MRPC dataset, where both the target model (including its embedding table) and the \langmodel{} are Llama-3.2-1B-Instruct with a vocabulary of size $128{,}256$ \cite{touvron2023llama}. 
The obfuscation is performed using isotropic Gaussian and Laplacian noise centered at zero, with the variance adjusted to different levels.

\approach{} consistently surpasses the nearest-neighbor baseline across all noise regimes.  In particular, at stringent privacy settings (low $\epsilon$, i.e. high noise), it delivers roughly 32\% improvement in token-recovery rate—demonstrating its robustness and making it a clear choice when privacy guarantees tighten. Furthermore, we show that the variance can be learned during training, and our decoding procedure compensates for stochasticity even at high variance. These findings are especially relevant given that prior work~\cite{mai2023split} applies local differential privacy (LDP) techniques to both input and output embeddings, typically comparing the obfuscated embeddings against a nearest-neighbor attack to demonstrate privacy strength. In contrast to the nearest-neighbor attack, \approach{} recovers a substantially higher fraction of tokens. Notably, expanding the beam size or increasing the candidate pool can further improve performance, albeit with additional computational costs.

\begin{figure}[t]
    \centering
    \includegraphics[width=0.6\linewidth]{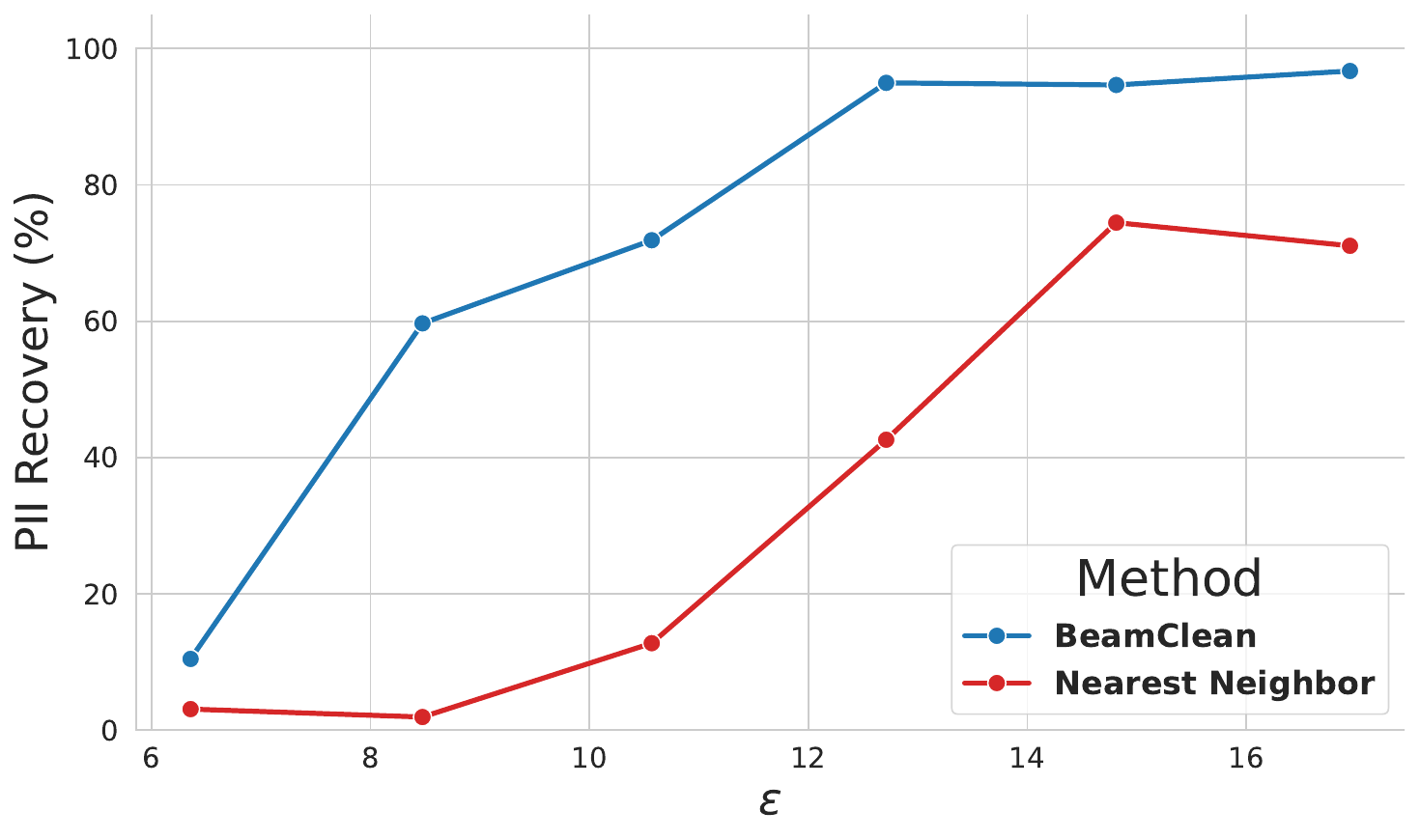}
    \caption{Mean PII Recovery percent on PAPILLON vs Laplace noise mechanism DP-$\epsilon$. \approach{} consistently able to recover more PII strings than Nearest Neighbor, recovering 60.0\% of PII strings compared to 1.9\% recovered by Nearest Neighbor at $\epsilon=8.5$.}
    \label{fig:pii-recovery}
\end{figure}

\paragraph{BeamClean recovers significantly higher PIIs compared to Nearest Neighbor.} 
\label{PII recovery}
Figure \ref{fig:pii-recovery} demonstrates the ability of \approach{} and Nearest Neighbor to recover PII strings from embeddings obfuscated with Laplacian noise mechanisms. For all the cases, \approach{} recovers more PII tokens than Nearest Neighbor. In particular, for an $\epsilon$ value of $8.5$ Nearest Neighbor can only recover 1.9\%, while \approach{} recovers 60\% of PII strings. This underscores the importance of having stronger privacy attacks to measure the obfuscation quality. In this situation, a practitioner may believe that they had protected almost all their PII data when in actuality less than a third would have been protected from \approach{}.

\label{subsub:static_vs_input_dep}
\begin{figure}[t]\centering\includegraphics[width=0.6\textwidth]{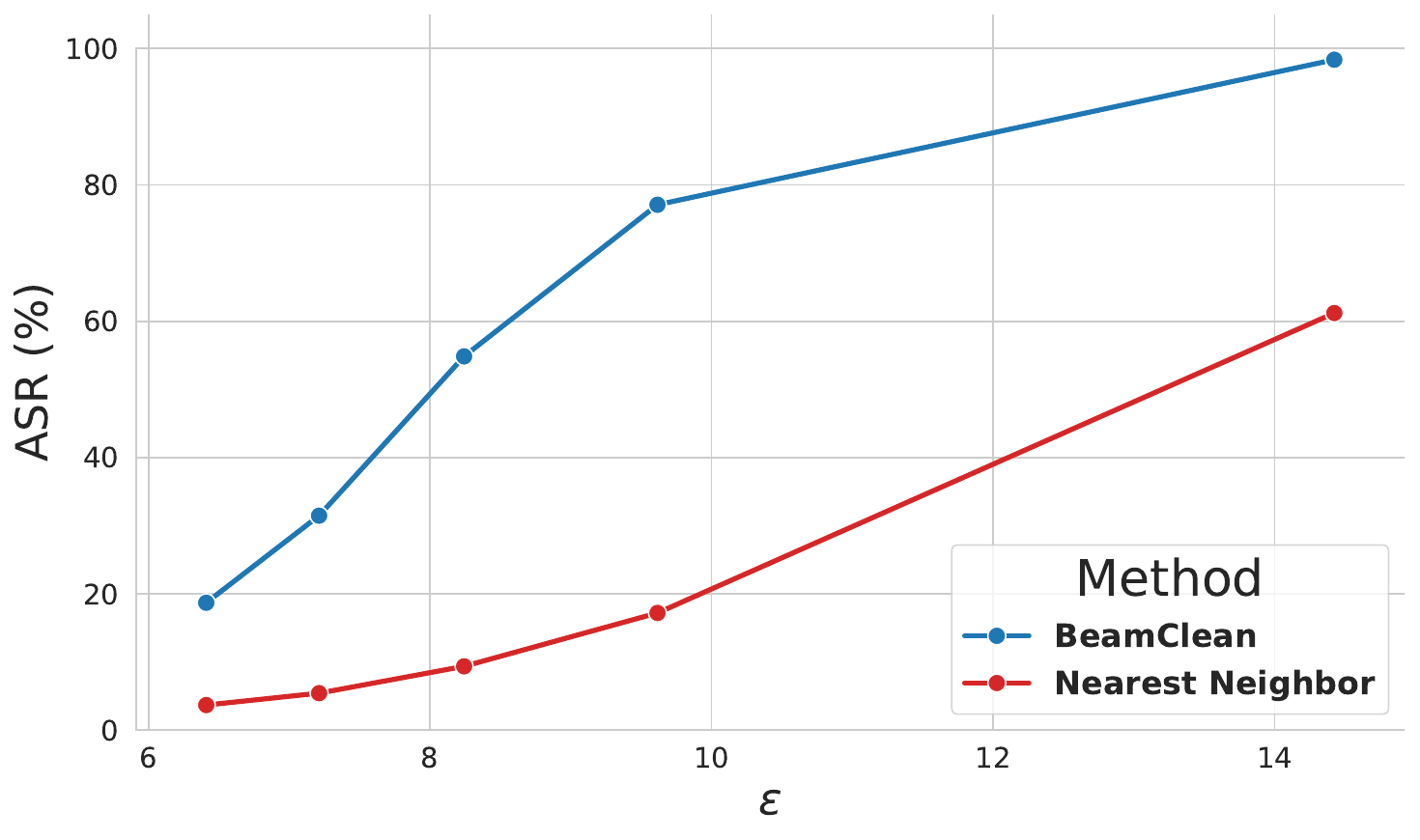}
\caption{ Attacking obfuscated GPT-2 embeddings using a Llama-3.2-1B-Instruct model as a language prior. \approach{} uniformly outperforms Nearest Neighbors, with the largest measured difference between the reconstruction methods occurring at DP $\epsilon \approx 9.6$, \approach{} achieves 77\% recovery versus 17\% for the baseline.}
\label{fig:llm-mismatch}
\end{figure}

\paragraph{The attacker does not need access to the target language model.}
\label{subsub:translation_res}
We evaluate a scenario where the target embeddings originate from a GPT-2 embedding table, yet the decoding prior is provided by a Llama-3.2-1B-Instruct model. Because these two models use different tokenization schemes, each GPT-2 token must be mapped or approximated to a corresponding LLaMA token. We build a direct $1{:}1$ mapping for each GPT-2 token to a unique LLaMA version of that token by choosing a restricted vocabulary, and we further restrict tokens to those present in MRPC. These experiments are run on the MRPC dataset from the GLUE benchmark.

Figure \ref{fig:llm-mismatch} shows the performance of \approach{} and Nearest Neighbor as a function of utility ($\epsilon$). We see that even when the language prior is different from the target model, \approach{} dominates Nearest Neighbor, showing ASR nearly $60\%$ higher compared to Nearest Neighbor when tested against Laplacian noise with a scale of $0.6$.

\section{Limitations}
\label{sec:limitations}
Although \approach{} demonstrates strong performance, several factors limit real-world applicability of this work. In our experimentation, we restricted ourselves to input independent noise mechanisms found in the literature. Though \approach{} is also applicable to more sophisticated input-dependent noise mechanisms, further experimentation is needed to determine its efficacy. From an algorithmic perspective, errors in decoding the earlier tokens in the sequences can propagate and escalate due to the autoregressive nature of language modeling. The beam strategy mitigates but does not eliminate, this risk. Further, large vocabularies and longer sequences greatly expand the search space, making exhaustive decoding expensive. Our beam-pruned approach helps, yet remains GPU-intensive. Despite these drawbacks, our results highlight that constant-noise mechanisms may be significantly more vulnerable to systematic attacks than prior work \cite{mai2023split} suggests.

\section{Conclusion \& Future Work}
\label{sec:conlusion}
We presented a novel attack framework for inverting obfuscated embeddings under a scenario wherein the adversary can only access leaked, noise-perturbed embeddings and the target language model’s embedding table. \approach{} combines learned noise-model estimation with language model priors in order to decode obfuscated embeddings to recover plaintext more effectively compared to naive nearest neighbor-based baselines used in input-independent Gaussian and Laplacian noise mechanisms. Moreover, we showed that \approach{} was able to consistently recover a significantly higher percentage of PII compared to the Nearest Neighbor attack. We also demonstrated that \approach{} does not necessarily need access to the target language model for the attack to be successful, inverting nearly $60\%$ more tokens than Nearest Neighbor when the decoding prior is a different model than the target language model. 

Future work could investigate how well \approach{} can reconstruct embeddings obfuscated with input \emph{dependent} noise. Additionally, varying hyperparameters, noise distributions, and model architectures further influence both obfuscation strength and attacker capabilities. Likewise, incorporating domain constraints, such as partial vocabulary knowledge or specialized language priors, offers promising directions for refining \approach{} and improving privacy-preserving designs in MaaS environments. 

\section{Ethics Statement}
\label{sec:ethics}
Our work introduces a state-of-the-art inversion of noise-perturbed embeddings back to plaintext. This plaintext could include prompts from users that could contain sensitive and proprietary information and so \approach{} represents an additional avenue of privacy leakage not previously addressed in the literature. We believe that the description and release of \approach{}, along with the associated source code and data, will enable security researchers to have a stronger method to measure the protections offered by their privacy enhancing technologies. 

\bibliographystyle{unsrtnat}  
\bibliography{references}

\clearpage

\appendix

\section{Inverse Problem} \subsection{Single Token}
\label{app:single_token} 

We start with a non-sequential (single-step) inverse problem where we observe:
\[
y = x + n, 
\quad 
n \sim \mathcal{M}(x; \mu(x), \Sigma(x)),
\]
and wish to infer \( x \in \mathbb{R}^d \). We have a prior \(p(x)\) (e.g., learned from some large dataset such as Alpaca).

Find parameters \(\theta\) that relate noisy data \( y \) to clean embeddings \( x \). A typical Bayesian formulation:
\begin{align}
    \hat{\theta} 
    &= \arg\max_\theta \log p(\theta \mid y) \\
    &= \arg\max_\theta \left[\log \int p(y \mid x, \theta)\,p(x)\,dx + \log p(\theta)\right].
\end{align}

\subsection{Sequential}
\label{app:sequential} 

\begin{align*}
    \hat{\theta}
    &= \arg\max_\theta \log p(\theta \mid \mathbf{y}_{1:T})  \\
    &= \arg\max_\theta \left[\log \int p(\mathbf{y}_{1:T} \mid \mathbf{x}_{1:T}, \theta)\,p(\mathbf{x}_{1:T})\,d\mathbf{x}_{1:T} + \log p(\theta)\right].
\end{align*}

\section{Inverse Problem (Token Sequence)}
We consider a sequence of noisy word embeddings \(\mathbf{y}_{1:T} = (\mathbf{y}_1, \mathbf{y}_2, \ldots, \mathbf{y}_T)\), where \(T\) is the sequence length. The corresponding clean embeddings (unobserved) are denoted as \(\mathbf{x}_{1:T} = (\mathbf{x}_1, \mathbf{x}_2, \ldots, \mathbf{x}_T)\).

The noise model assumes that the noisy embedding \(\mathbf{y}_t\) at time \(t\) depends on all previous clean embeddings \(\mathbf{x}_{1:t}\), parameterized by \(\theta\):
\[
p(\mathbf{y}_t \mid \mathbf{x}_{1:t}, \theta).
\]

The prior on the clean embeddings \(\mathbf{x}_{1:T}\) is given by a pretrained language model:
\[
p(\mathbf{x}_{1:T}) = p(\mathbf{x}_1) \prod_{t=2}^T p(\mathbf{x}_t \mid \mathbf{x}_{<t}).
\]

The goal is to maximize the marginal likelihood of the noisy embeddings:
\[
\log p(\mathbf{y}_{1:T} \mid \theta) = \log \int p(\mathbf{y}_{1:T}, \mathbf{x}_{1:T} \mid \theta) \, d\mathbf{x}_{1:T}.
\]

\section{Additive‑Noise Mechanisms for Differential Privacy}
\label{app:epsilon_calc}

\label{app:additive-noise}
\subsection*{1 \; Global Sensitivity}
For a real‑valued query \(f:\mathcal{D}\!\to\!\mathbb{R}\), the (global) \(\ell_p\)-sensitivity is
\begin{equation}
  \Delta_p f \;=\;
  \max_{\substack{x,y\in\mathcal{D}\\\|x-y\|_0\le 1}}
  \bigl\|f(x)-f(y)\bigr\|_{p},
\label{eq:sensitivity}
\end{equation}
i.e.\ the greatest change in the output when one record is added or removed.
Equations~\eqref{eq:laplace}--\eqref{eq:vector} calibrate noise in terms of
\(\Delta_1 f\) or \(\Delta_2 f\).

\subsection*{2 \; Laplace Mechanism (\(\varepsilon\)-DP)\,\cite{dwork2006calibrating}}
\begin{equation}
  \boxed{
  \mathcal{M}_{\text{Lap}}(x;\,f,\varepsilon)
  \;=\;
  f(x)
  \;+\;
  \mathrm{Laplace}\!\bigl(0,\;b=\tfrac{\Delta_1 f}{\varepsilon}\bigr)
  }
\label{eq:laplace}
\end{equation}
Adding \emph{i.i.d.} Laplace noise with scale
\(b=\Delta_1 f/\varepsilon\) guarantees \(\varepsilon\)-differential privacy since
\[
  \frac{\Pr[\mathcal{M}_{\text{Lap}}(x)=z]}
       {\Pr[\mathcal{M}_{\text{Lap}}(y)=z]}
  \;\le\; e^{\varepsilon},
  \quad
  \forall\;z,\;x\!\sim\!y.
\]

\textbf{Discrete variant.}  Replacing continuous Laplace noise by the
\emph{geometric} (discrete Laplace) distribution yields the universally
utility‑maximizing geometric mechanism\,\cite{ghosh2009universally, gupte2010universally}.

\subsection*{3 \; Gaussian Mechanism (\(\varepsilon,\delta\)-DP)\,\cite{dwork2014algorithmic}}
\begin{equation}
  \boxed{
  \mathcal{M}_{\text{Gauss}}(x;\,f,\varepsilon,\delta)
  \;=\;
  f(x)
  +
  \mathcal{N}\!\Bigl(
     0,\;
     \sigma^{2}
     =\frac{2\,\ln\!\bigl(1.25/\delta\bigr)\,(\Delta_2 f)^2}{\varepsilon^{2}}
  \Bigr)
  }
\label{eq:gauss}
\end{equation}
With \(0<\varepsilon<1\) and \(0<\delta<1\), the variance choice above
ensures (\(\varepsilon,\delta\))-differential privacy by bounding the overlap
between the two Gaussian output distributions corresponding to neighboring datasets.

\subsection*{4 \; Vector‑Valued Queries (\(d>1\))}
For \(f:\mathcal{D}\!\to\!\mathbb{R}^{d}\) (\(d\ge 2\)), add independent noise
per coordinate:
\begin{equation}
  \boxed{
  \mathcal{M}_{\text{Gauss}}^{d}(x)
  \;=\;
  f(x)+(\eta_1,\dots,\eta_d),
  \qquad
  \eta_i\overset{\text{i.i.d.}}{\sim}\mathcal{N}\!\bigl(0,\sigma^2\bigr)
  }
\label{eq:vector}
\end{equation}
where \(\sigma^2\) is given by Equation~\eqref{eq:gauss} and the sensitivity
\(\Delta_2 f\) is computed in the \(\ell_2\) norm.

\end{document}